\documentclass[journal,lettersize]{IEEEtran}
\usepackage{amsmath,amsfonts} 
\usepackage{array}
\usepackage[caption=false,font=normalsize,labelfont=sf,textfont=sf]{subfig}
\usepackage{algorithmic}
\usepackage{algorithm}
\usepackage{multirow}
\usepackage{textcomp}
\usepackage{stfloats}
\usepackage{url}
\usepackage{tabularx}
\usepackage{verbatim}
\usepackage{dsfont}
\usepackage{bm}
\usepackage{graphicx}
\usepackage{tabularray}
\usepackage[table,xcdraw,dvipsnames]{xcolor}
\usepackage{amsmath}
\usepackage{hyperref}
\usepackage{cite}
\usepackage{booktabs}
\usepackage{colortbl}

\usepackage{enumitem}

\algsetup{ linenosize=\small, linenodelimiter=. }
\usepackage{tikz}
\usetikzlibrary{matrix, positioning, patterns, shapes, arrows}
\usepackage{pgfplots}
\newcommand{\argmin}{\mathop{\mathrm{argmin}}\limits} 
\usepgfplotslibrary{groupplots}
\usepgfplotslibrary{colorbrewer}
\usetikzlibrary{spy,backgrounds}
\usepackage{epsfig}
\usepackage{pgfplots}
\pgfplotsset{compat=newest}
\usepgfplotslibrary{groupplots}
\usepackage{tikz}
\usetikzlibrary{matrix, positioning, patterns, shapes, arrows}
\usepgfplotslibrary{colorbrewer}
\pgfplotsset{compat = 1.15, cycle list/Set1-8} 
\usetikzlibrary{pgfplots.statistics, pgfplots.colorbrewer} 

\usepackage{pgfplotstable}

\definecolor{Paired-1}{RGB}{31,120,180}
\definecolor{Paired-2}{RGB}{166,206,227}
\definecolor{Paired-3}{RGB}{51,160,44}
\definecolor{Paired-4}{RGB}{178,223,138}
\definecolor{Paired-5}{RGB}{227,26,28}
\definecolor{Paired-6}{RGB}{251,154,153}
\definecolor{Paired-7}{RGB}{255,127,0}
\definecolor{Paired-8}{RGB}{253,191,111}
\definecolor{Paired-9}{RGB}{106,61,154}
\definecolor{Paired-10}{RGB}{202,178,214}
\definecolor{Paired-11}{RGB}{177,89,40}
\definecolor{Paired-12}{RGB}{105,105,105}
\definecolor{Paired-13}{RGB}{80,80,80}

\definecolor{Color1}{HTML}{D81B60}
\definecolor{Color2}{HTML}{1E88E5}
\definecolor{Color3}{HTML}{FFC107}
\definecolor{Color4}{HTML}{004D40}

\newcommand*\pcfrozen{$PC_{Frozen}$ }
\pgfplotsset{compat=1.18}

\usepackage{amsmath,amsthm}
\newtheorem{theorem}{Theorem}

\newtheorem{corollaryinner}{Corollary}[theorem] 
\newenvironment{corollary}[1][]
 {
  \if\relax\detokenize{#1}\relax
  \else
    \ifcsname #1-used\endcsname
      \expandafter\xdef\csname #1-used\endcsname{\the\numexpr\csname #1-used\endcsname+1}%
    \else
      \expandafter\gdef\csname #1-used\endcsname{1}%
    \fi
    \renewcommand{\thecorollaryinner}{\ref{#1}.\csname #1-used\endcsname}%
  \fi
  \corollaryinner
 }
 {\endcorollaryinner}

\begin{document}

\title{Enabling Fast Polar SC Decoding with IR-HARQ}
\author{Marwan Jalaleddine, Mohamad Ali Jarkas, Jiajie Li, Warren J. Gross
\thanks{ Marwan Jalaleddine, Mohamad Ali Jarkas, Jiajie Li and Warren J. Gross are with the Department of Electrical and Computer Engineering at McGill university, Montreal, Quebec, Canada.  Their emails are: marwan.jalaleddine@mail.mcgill.ca,mohamad.jarkas@mail.mcgill.ca, jiajie.li@mail.mcgill.ca,   warren.gross@mcgill.ca.}
}



\maketitle

\begin{abstract}
To extend the applications of polar codes within next-generation wireless communication systems, it is essential to incorporate support for Incremental Redundancy (IR) Hybrid Automatic Repeat Request (HARQ) schemes. For very high-throughput applications, Successive Cancellation (SC) decoding is particularly appealing for polar codes owing to its high area efficiency. In this paper, we propose modifications to SC decoders that employ special nodes to accelerate decoding. Our modifications enable the use of polar IR-HARQ with SC decoding for high throughput applications. Compared to the unmodified SC IR-HARQ scheme, our proposed approach allows us to achieve up to {$66 \%$ reduction in node traversals with polar codes}. Simulation results confirm that the proposed special node modifications do not cause any degradation in FER performance { compared to the unmodified state-of-the art special nodes}.
\end{abstract}


\section{Introduction}
Polar codes are a capacity-achieving error-correcting code \cite{arikan_channel_2009} that has been incorporated into the 5G control channel \cite{3gpp_nr_2021}. More recently, there has been interest in incorporating polar codes in the data channel for the 6G standard \cite{Huaweiharq}. This necessitates support for polar { Hybrid Automatic Repeat Request (HARQ)} \cite{Huaweiharq}, which aims to enhance link reliability through re-transmissions.

HARQ schemes are typically categorized into two main categories: Chase Combining (CC) HARQ and Incremental Redundancy (IR) HARQ. In CC-HARQ, the same coded block is retransmitted across multiple transmissions, and this method effectively boosts received signal power. Conversely, IR-HARQ improves performance by transmitting new redundancy bits with each retransmission.  {Although CC-HARQ is simpler to implement in practice, IR HARQ generally outperforms CC-HARQ in terms of error correction performance \cite{cheng_coding_2006}}. 

CC-HARQ schemes for polar codes were initially introduced in \cite{chen_polar_2014}, where selective puncturing and re-transmissions of bits were employed based on greedy search algorithms. However, these methods require meticulous design of puncturing patterns, as such patterns critically influence the capacity of the virtual sub-channels assigned to message bits. Moreover, these approaches exhibit notable degradation in error-correction performance, primarily due to the Chase combining of log-likelihood ratios (LLRs) \cite{cheng_coding_2006,zhao_adaptive_2018}.

To mitigate the degradation in decoding performance observed in CC-HARQ schemes, \cite{li_capacity-achieving_2016} introduced a polar IR-HARQ scheme employing an incremental freezing mechanism; however, since the two transmissions are decoded independently as short codes, this scheme fails to realize the coding gains typically associated with longer code lengths \cite{arikan_channel_2009}. This method was later enhanced by \cite{zhao_adaptive_2018}, who proposed concatenating multiple transmissions into a single, longer polar code to improve both coding and diversity gains. This enhancement was realized through the incremental extension of the polar encoding matrix, thereby exploiting the superior decoding performance characteristic of longer polar codes { and achieving a $1$ dB gain in performance compared to its CC-HARQ equivalent at a target FER of $10^{-2}$ dB}.

Despite the advancements made by \cite{zhao_adaptive_2018}, the newly introduced bit types restrict the algorithm's use of {specialized decoders referred to as special nodes. These special nodes allow decoding of a group of bits in parallel if they have a special structure. Without the use of special nodes, the SC and SCL decoder have to decode each bit sequentially which introduces additional latency.} 

In \cite{jalaleddine_hardware-friendly_2025} the authors devised a method capable of integrating special nodes with Successive Cancellation List (SCL) decoding. This was done by modifying the candidate generation method by representing all possible special nodes as linear combinations of the outputs of the basic special node types and the values of a newly introduced bit type. The same binary vector-based bit type generation method can be used with SC decoding; however, applying the same candidate generation method on SC decoding would introduce high computational complexity overhead as SC decoding only chooses the most likely candidate for that node and does not explore all possible candidates. Hence, a completely different technique should be employed to allow for IR-HARQ enabled fast SC decoding.

This work introduces modifications to the state-of-the-art SC special node algorithms \cite{10293973,sarkis_fast_2014} to enable their integration with polar IR-HARQ in SC decoding. Unlike the method proposed for SCL decoding \cite{jalaleddine_hardware-friendly_2025}, the method introduced here is optimized for SC decoding with little complexity overhead compared to the existing SC decoding special node algorithms. The incorporation of special nodes in the IR-HARQ framework effectively reduces the number of node traversals required during SC decoding by up-to {$66\%$} with polar codes, thereby lowering the overall decoding latency. 

\section{Preliminaries}
\subsection{Notation}
Matrices are denoted by a bold upper-case letter ($\bm{M}$), while vectors are denoted by bold lower-case letters ($\bm{v}$). The $i^{\text{th}}$ element of a vector $\bm{v}$ is denoted as $v_i$. The vector elements are numbered from left to right starting from an index of 0. Operation $\oplus$ represents XOR binary bit-wise operation. Furthermore, this work considers $(n,k)$ linear block codes, where $n$ is the code length and $k$ is the code dimension. The  modulo operator is represented by the function$\text{ mod } $, and the Boolean complement of $\rho$ is denoted as $\bar{\rho}$. The sign function is defined as:
\[
    sign(\zeta) = 
    \begin{cases}
        0, & \zeta \ge 0, \\
        1, & \zeta < 0.
    \end{cases}
\] 
\vspace{-0.8cm}
\subsection{Polar Codes}
\label{sec:PC}

Encoding and decoding of a polar code can be described using a binary tree representation with multiple stages $s$. We refer to each polar sub-code in a polar tree as a node with size $N_v = 2^s$ and to nodes at the lowest stage (stage 0) as leaf nodes. First, before polar encoding can happen, the $k$ most reliable bit-channels are determined. Frozen bits are the $n-k$ least reliable bits that are set to a predefined value (typically $0$) which is known to the decoder. The encoding process ascends the binary tree applying the equation $\langle u_{j:j\times2^{s-1}} \oplus u_{j\times2^{s-1}:2^s-1}, u_{j\times2^{s-1}:2^s-1} \rangle$ for all $j\in [0,\frac{n}{2^s}-1]$ and for stages $1 \leq s\leq log_2(n) $ where at each stage $s$ in the tree $\frac{n}{2^s}$ new polar codes of length $2^s$ are generated in parallel. Due to the fact that the inverse of this encoding operation is the same as the encoding procedure \cite{arikan_channel_2009}, retrieving the message bits can be done by following the same process as encoding. The polar decoder on the other hand receives $\mathbf{y}$ as its noisy channel output using which the LLRs can be calculated. SC decoding proceeds to descend the tree to stage 0 and estimate the resulting bit values from both the LLRs and the previously estimated bits at each stage \cite{arikan_channel_2009}. 
\subsection{Special Nodes}
Instead of traversing the entire tree down to stage $0$ with SC decoding, special sub-trees/nodes of the polar decoding tree can be decoded efficiently if specific patterns of the frozen and information bits are encountered \cite{sarkis_fast_2014} \cite{10293973}. 
The main special nodes that we will consider are the Rate 0, Rate 1,  SPC and REP established in \cite{sarkis_fast_2014}, and SPC-2, REP-2, PCR and RPC introduced in \cite{10293973}. We provide a summary of the structure and decoding algorithms introduced in \cite{sarkis_fast_2014} and \cite{10293973} below.
\subsubsection{Rate 0 Node}
A Rate 0 node is defined as a node in which all leaf node bits are frozen. In this case, the decoding output is the all zero codeword.
\subsubsection{Rate 1 Node}
A Rate 1 node is defined as a node where all the leaf node bits are information bits. In this case, the output codeword bits for this node are found by obtaining the hard decision of the node LLRs.
The hard decision of $\bm{l}$, is defined as $\bm{HD}$, given by $sign(\bm{l})$.

\subsubsection{REP Node}
A repetition (REP) node is defined as a node where all the leaf bits are frozen, except for the right-most bit. After encoding, this results in a codeword where the information bit from the right-most position is repeated in every bit bit position of the node. In this case, decoding can proceed by summing all the LLRs of the node and using a $sign$ function to determine the value of the repeated bit.

\subsubsection{REP-2 Node}
A REP-2 node is defined as a node in which all leaf bits are frozen, except for the rightmost two bits. This configuration results in the formation of two repetition codes. The first repetition code corresponds to the repetition of a bit value on even-numbered indices of the node, while the second repetition code is associated with a repetition of a bit value on the odd-numbered indices of the node. Hence, two REP decoders can be used to determine the value of the repeated bits in parallel, one REP decoder used on the even indices and one REP decoder used on the odd indices.

\subsubsection{PCR Node}
A Parity Checked Repetition (PCR) node is characterized as a node where the three rightmost leaf bits are information bits, while the remaining bits are all frozen bits. The codeword is divided into four groups $X^z$. Each set $X^z$ contains all bit positions whose index $j$ satisfies the following condition: $j \text{ mod }4 =z$. The structure of this node guarantees that $X^0$ is the repetition of $0$, $X^1$ is the repetition of the $N_v -3$ codeword bit, $x_{N_v -3}$,  $X^2$ is the repetition of the $N_v -2$ codeword bit, $x_{N_v -2}$ and $X^3$ is the repetition of the $N_v -1$ codeword bit $x_{N_v -1}$. 
\subsubsection{SPC Node}
A SPC (Single Parity Check) node is defined as a node where only the leftmost leaf bit is frozen, with the remaining bits being information bits. 

The frozen bit $p_0 = 0$ enforces an even parity constraint. To decode a regular SPC node, we first take the hard decision on the node LLRs. The parity of the bits is then calculated. If we have an even parity, the parity constraint is satisfied, and $\bm{HD}$ is returned as the codeword result. If we have an odd parity, the bit corresponding to the lowest LLR magnitude, $\argmin(|l_j|),$ is flipped in the returned $\bm{HD}$ codeword result.

\subsubsection{SPC-2 Node}
A SPC-2 node is defined when the leftmost two leaf bits are frozen and all the the remaining bits are information bits. This leads to the creation of two parity constraints, one on the even bits and one on the odd bits. Two SPC decoders can then be used to decode odd and even indices respectively.

\subsubsection{RPC Node}
An RPC (Repeated Parity Check) node is defined as a node where the leftmost three leaf bits are frozen, while the remaining bits are information bits. Using the notation $C^z=\bigoplus x \in X^z , z \in \{0,1,2,3\}$, the parity equations of this node are:
\begin{equation}
     \begin{split}    C^1 \oplus C^3=
        C^{2}  \oplus  C^3 =
         & C^1 \oplus C^2  = 
         C^0  \oplus C^3 =0 \text{.} \label{eq:org_rpc2}
     \end{split}
\end{equation}
 We then proceed to decode each group through a similar procedure to the SPC node \cite{10293973}.

\subsection{Polar IR-HARQ by Matrix Extension}

When employing IR-HARQ, the receiver requests additional redundancy from the transmitter to enhance decoding performance. In the context of polar codes, the IR-HARQ by matrix extension approach enlarges the polar encoding tree to accommodate an extended codeword length, equal to the sum of the original code length and the newly transmitted redundancy. Subsequently, the newly introduced bit-channels may be selected as information bits if their reliability surpasses that of previously transmitted information bits. In such cases, the original information bits are reclassified as Parity-Check frozen (\pcfrozen) bits, with their values mapped one-to-one to the corresponding new information bits. During encoding, each \pcfrozen bit shares the same value as its mapped information bit. During decoding, however, the new information bit is first estimated, after which the corresponding \pcfrozen bit is fixed to that decoded value. Thus, \pcfrozen bits constitute a special category of frozen bits that can assume binary values of either $0$ or $1$, with their values determined post-decoding of their associated information bits. Since conventional special node decoding algorithms \cite{10293973,sarkis_fast_2014} typically assume frozen bits to be 0, these decoding algorithms need to be modified.
\subsection{IR-HARQ enabled SCL Special Nodes}

For SCL decoding, \cite{jalaleddine_hardware-friendly_2025} proved that the resulting candidates $\bm{\beta}$ of the special node in the presence of \pcfrozen bits is the linear combination of two vectors, the original candidates $\bm{\beta'}$ (assuming there are no \pcfrozen bits) and the encoded \pcfrozen bits $\bm{pc}$ such that:
\begin{equation}
 \bm{\beta} = \bm{\beta'}  \oplus \bm{pc} .
\end{equation}

After generating all possible candidates for a node, the SCL algorithm then chooses a certain number of most likely candidates to proceed with. For SC decoding, only the most likely candidate is chosen at all times through using optimized algorithms; hence, applying the same method to implement fast nodes for IR-HARQ as that with SCL decoding results in additional unnecessary computations. This means that computationally efficient modifications to the SC special nodes are needed for IR-HARQ enabled SC decoding.

\section{IR-HARQ enabled SC Special Nodes}
This section presents modifications to extend SC special node decoding capabilities to support frozen bits with arbitrary binary values, such as those encountered with IR-HARQ.

\subsection{Rate 0 and Rate 1 nodes}
The decoding of rate 0 and rate 1 nodes is trivial. For rate 0 nodes, the result of the decoding is simply the encoded \pcfrozen bits similar to \cite{jalaleddine_hardware-friendly_2025}. For the rate 1 node, the result of the decoder is the hard decision of the the received signal \cite{alamdar-yazdi_simplified_2011}, since it does not have any \pcfrozen bits.

\subsection{REP Node}
\begin{theorem} The repeated bit in a repetition node can be determined through the following calculation:
\begin{equation}
    i = sign(\sum_{j=0}^{N_v -1} l_j \, (1 - 2\,pc_j)),
\end{equation}
where $l_j$ denotes the LLR corresponding to the $j$-th bit.
The corresponding codeword is then generated by computing:
\begin{equation}
\beta_j = i  \oplus pc_j, \quad \forall j\in[0,N_v-1].
\end{equation}

\end{theorem}
\begin{proof}
The most likely codeword corresponds to the codeword with the minimum weighted hamming distance \cite{9613322} which is equivalent to the path metric (PM). For a repetition node, there are only two candidate paths, corresponding to information bit values $0$ and $1$, with path metrics defined as
\begin{align}
    PM^0 &= \sum_{j=0}^{N_v -1} |l_j| \, \big( (pc_j \oplus 0) \oplus HD_j \big), \\
    PM^1 &= \sum_{j=0}^{N_v -1} |l_j| \, \big( (pc_j \oplus 1) \oplus HD_j \big),
\end{align}
where $HD_j$ denotes the hard decision on $l_j$.

The ML decision corresponds to the path with the smaller path metric.  
Therefore, path-1 is chosen if $PM^1 \leq PM^0$, i.e.:
\begin{equation}
    PM^1 - PM^0 = \sum_{j=0}^{N_v -1}\kappa_j < 0, \label{eq:rep-1}
    \vspace{-0.3cm}
\end{equation} 

else path-0 is chosen where,
\begin{equation}
    \kappa_j= |l_j| \Big[ (\bar{pc_j} \oplus HD_j) - (pc_j \oplus HD_j) \Big] .
\end{equation}

We now analyze the sign of this difference term by considering both cases for $pc_j$. 

\begin{itemize}
    \item If $pc_j = 1$, then $\bar{pc_j} = 0$, and the expression reduces to
    \begin{equation}
         \kappa_j =  |l_j| \big(HD_j - \bar{HD_j}\big)
        = - l_j.
    \end{equation}
    \item If $pc_j = 0$, then $\bar{pc_j} = 1$, yielding
    \begin{equation}
      \kappa_j  =  |l_j| \big(\bar{HD_j} - HD_j\big)
        =  l_j.
     \end{equation}
\end{itemize}

Combining both cases, (\ref{eq:rep-1}) can be expressed compactly as:
\begin{equation}
\vspace{-0.1cm}
    PM^1 - PM^0 = \sum_{j=0}^{N_v -1} l_j \, (1 - 2\,pc_j),
\end{equation}
If $PM^1 - PM^0 <0$ , $i=1$ while if $PM^1 - PM^0 \geq0$ , $i=0$. Hence determining $i$ is equivalent to using the sign function on $PM^1-PM^0$. 
\end{proof}
\subsection{REP-2 Node}
\begin{corollary}
The information bits in a REP-2 Node can be decoded using :
\vspace{-0.4cm}
\begin{align}
        i_{N_v -1}&=sign(\sum_{j=0}^{\frac{N_v}{2}-1} l_{2j+1} \, (1 - 2\,pc_{2j+1})), \\
    i_{N_v -2}&=sign(\sum_{j=0}^{\frac{N_v}{2}-1} l_{2j} \, (1 - 2\,pc_{2j})) .
    \end{align}
\end{corollary}
The corresponding codeword is the repetition of $i_{N_v -1} $ on the odd bits and the repetition of $i_{N_v -2}$ on the even bits XORed with the \pcfrozen encoded vector. For $j \in [0,\frac{n}{2}-1]$ this can be represented as:
\begin{align}
 \beta_{2j} = i_{N_v -2}  \oplus pc_{2j}, \text{ and }
 \beta_{2j+1} = i_{N_v -1}  \oplus pc_{2j+1}.
\end{align}
\begin{proof}
    A REP 2 node can be considered as two REP nodes, one on the even  bits which represents the repetition of the encoded $i_{N_v -2}$ and one on the odd bits which corresponds to the repetition of the encoded $i_{N_v -1}$. Hence, since the even and odd bits can be decoded independently, two modified REP nodes can be used. 
\end{proof}
\subsection{PCR node} 
\begin{corollary}
 We can apply the same decoding method as the modified repetition node in Theorem 1 to determine the individual bits corresponding to $X^0$, $X^1$, $X^2$, $X^3$. The necessary modifications are:
\begin{enumerate}
    \item Modifying the calculation of path metric depending on the value of $pc_j$: \[
    PM =  PM +(1-2\times pc_j)\times l_j.
\]
\item XOR-ing the result of the PCR node $\hat{\bm{x}}$ with the encoded \pcfrozen vector $\bm{pc}$.
\end{enumerate}
\end{corollary}
\begin{proof}
    Since this node corresponds to 4 different repetition nodes, the decoding procedure of the 4 repetition nodes can proceed as described before.
\end{proof}
\subsection{SPC Node}
\begin{theorem}  
The SPC node decoding can be performed by following the following a modified Wagner decoder:
through flipping bit index $\argmin_{j} \left| l_j \right |$ if:
    \begin{equation}
        (\bigoplus_{j=0}^{N_v -1} HD_j)\oplus pc_0 \neq 0.
    \end{equation}
    \vspace{-0.6cm}
\end{theorem}
\begin{proof}
 Assume we have a vector composed of ${HD}_0, {HD}_1, ... {HD}_{N_v-1}$ at stage s. By descending the tree to stage 0, the leftmost bit, $p_0$ corresponds to:  
 \begin{equation}
     p_0 = \bigoplus_{j=0}^{N_v -1} HD_j.
 \end{equation}
This is equivalent to a parity check constraint applied on the codeword of : \begin{equation}
    \left(\bigoplus_{j=0}^{N_v -1} HD_j \right)\oplus p_0 =0
\end{equation} 
The left-most bit is a special case where the encoded \pcfrozen bit equals to the \pcfrozen bit at stage 0, $pc_0=p_0$.
Hence, flipping the least reliable bit only occurs if the new $ \left(\bigoplus_{j=0}^{N_v -1} HD_j \right)\oplus pc_0$ is violated.
\end{proof}
\subsection{SPC-2 Node}
\begin{corollary}
Decoding an SPC-2 node can be done by using two SPC nodes, one on the even bits and one on the odd bits. 
\begin{enumerate}
    \item Flip bit index $2\times \argmin_{j} \left| l_{2j }\right|$ if:
    \begin{equation}
(\bigoplus_{j=0}^{\frac{N_v}{2} -1} HD_{2j})\oplus pc_0 \neq 0.
    \end{equation}
    \item Flip bit index $2\times \argmin_{j} \left| l_{2j+1}\right| +1$ if :
    \begin{equation}
(\bigoplus_{j=0}^{\frac{N_v}{2} -1} HD_{2j+1})\oplus pc_1 \neq 0. \end{equation}
\end{enumerate}

\end{corollary}
\begin{proof}
The parity check equations for the SPC 2 node are:
\begin{align}
p_0 = \bigoplus_{j=0}^{N_v -1} HD_j,\quad \text{ and } \quad
p_1 = \bigoplus_{j=0}^{\frac{N_v}{2} -1} HD_{2j+1}.
\end{align}

where $p_j$ is the $j^{th}$ parity bits at stage $0$.

The encoded \pcfrozen bits ($pc_0 = p_0 \oplus p_1$, $pc_1 = p_1$) are:
\begin{align}
      pc_0 =  \bigoplus_{j=0}^{\frac{N_v}{2} -1} HD_{2j}, \quad \text{ and } \quad
      pc_1 = \bigoplus_{j=0}^{\frac{N_v}{2} -1} HD_{2j+1}.
\end{align}
Hence, the same procedure taken with an SPC node can be used in this case, using two SPC modules. 
\end{proof}
\subsection{RPC Node}
\begin{corollary}
A modification for the RPC algorithm to support decoding these new equations would be XOR-ing the $C^z$ with $\{pc_0 , pc_1 , pc_2, 0\}$. 
\begin{proof}
For the RPC node, the new parity equations with the presence of \pcfrozen bits are:
\begin{equation}
    \begin{split}
    {\color{blue}p_1 \oplus C^1} \oplus C^3 &=
         {\color{orange} C^2 \oplus p_2}  \oplus C^3\\
         &=
         {\color{blue}C^1 \oplus p_1 }\oplus {\color{orange}C^2  \oplus p_2}\\
         &= 
        {\color{teal} C^0 \oplus p_0 \oplus p_1 \oplus p_2  } \oplus C^3 =0\text{.}
    \end{split} \label{eq:rpc1}
\end{equation}
\begin{figure}[!t]
  \centering
  \begin{tikzpicture}[spy using outlines = {rectangle, magnification=2.0, connect spies}]
    \begin{groupplot}[group style={group name=fer_queries, group size= 2 by 1, horizontal sep=5pt, vertical sep=5pt},
      footnotesize,
      height=.6\columnwidth,  width=0.80\columnwidth,  
      xlabel=$\frac{E_s}{N_0}$ (dB),
      ymode=log,
      tick align=inside,
      grid=both, grid style={gray!30},
      /pgfplots/table/ignore chars={|},xtick={-9,-8,...,1},
      ] 
      \nextgroupplot[ylabel= FER, ytick pos=left, y label style={at={(axis description cs:-0.225,.5)},anchor=south},ymin=1e-4, ymax = 1, xmin=-9, xmax=-2]
\addplot[mark=none, Paired-3 , semithick]  table[x=Es/N0, y=FER] {plots/sc/HARQ_special/2048_f_3072.txt};\label{gp:3072}
\addplot[mark=none, Paired-5, semithick]  table[x=Es/N0, y=FER] {plots/sc/HARQ_special/2048_f_4096.txt};\label{gp:4096}
\addplot[mark=none, Paired-7, semithick]  table[x=Es/N0, y=FER] {plots/sc/HARQ_special/2048_f_5120.txt};\label{gp:5120}
\addplot[mark=none, Paired-9, semithick]  table[x=Es/N0, y=FER] {plots/sc/HARQ_special/2048_f_6144.txt};\label{gp:6144}
\addplot[mark=none, Paired-11, semithick]  table[x=Es/N0, y=FER] {plots/sc/HARQ_special/2048_f_7168.txt};\label{gp:7168}
\addplot[mark=none, Paired-13,semithick]  table[x=Es/N0, y=FER] {plots/sc/HARQ_special/2048_f_8192.txt};\label{gp:8192}

\addplot[mark=none, Paired-3 , dashed]  table[x=Es/N0, y=FER] {plots/sc/HARQ_no_special/2048_f_3072ns.txt};
\addplot[mark=none, Paired-5, dashed]  table[x=Es/N0, y=FER] {plots/sc/HARQ_no_special/2048_f_4096ns.txt};
\addplot[mark=none, Paired-7, dashed]  table[x=Es/N0, y=FER] {plots/sc/HARQ_no_special/2048_f_5120ns.txt};
\addplot[mark=none, Paired-9, dashed]  table[x=Es/N0, y=FER] {plots/sc/HARQ_no_special/2048_f_6144ns.txt};
\addplot[mark=none, Paired-11, dashed]  table[x=Es/N0, y=FER] {plots/sc/HARQ_no_special/2048_f_7168ns.txt};
\addplot[mark=none, Paired-13, dashed]  table[x=Es/N0, y=FER] {plots/sc/HARQ_no_special/2048_f_8192ns.txt};

      \coordinate (top) at (rel axis cs:0,1);
      \coordinate (spypoint1) at (axis cs:7.45,2e-7);
      \coordinate (magnifyglass1) at (axis cs:2.6,1.1e-5);
      \coordinate (bot) at (rel axis cs:1,0);
    \end{groupplot}
    \path (top|-current bounding box.north) -- coordinate(legendpos) (bot|-current bounding box.north);
    \matrix[
    matrix of nodes,
    anchor=south,
    draw,
    inner sep=0.2em,
    draw
    ]at(legendpos) 
    {
    \ref{gp:3072}& \tiny  3072 &[1pt] 
    \ref{gp:4096}& \tiny  4096 &[1pt]
    \ref{gp:5120}& \tiny  5120&[1pt] \\
    \ref{gp:6144}& \tiny  6144 &[1pt] 
    \ref{gp:7168}& \tiny  7168 &[1pt]  
    \ref{gp:8192}& \tiny  8192\\
      };
  \end{tikzpicture}
  \vspace*{-0.5cm}
  \caption{\label{fig:FER_2} Comparison of decoding performance for polar code configurations with feedback. Straight lines denote the proposed special nodes; dashed lines represent previous works.}
   \vspace*{-0.5cm}
\end{figure}
We can note that that the encoded \pcfrozen bits can be written as $pc_0 =p_0 \oplus p_1 \oplus p_2  $, $pc_1 =p_1 $ and $pc_2=p_2$. Additionally, we can use the change of variables for the colored variables in (\ref{eq:rpc1}) such that: ${ C^0 \oplus pc_0  } = \widetilde{C}^0 $, ${ C^1 \oplus pc_1  } = \widetilde{C}^1 $, and ${ C^2 \oplus pc_2  } = \widetilde{C}^2 $.

Hence, equation~\eqref{eq:rpc1} can be simplified to
\begin{subequations}
\begin{align}
         \widetilde{C}^1 \oplus C^3= 
        \widetilde{C}^{2}  \oplus C^3  =
         {\widetilde{C}^1}&\oplus \widetilde{C}^2  = 
        \widetilde{ C}^0  \oplus C^3 =0\text{.}
     \label{eq:rpc4}
     \end{align}
\end{subequations}

After changing variables, (\ref{eq:rpc4}) is identical to (\ref{eq:org_rpc2}).
\end{proof}
\end{corollary}
\begin{table*}[htbp]
\centering
\caption{The number of node traversals with IR HARQ feedback for polar codes.}
\label{tab:performance}
\resizebox{!}{2.5cm}{
\renewcommand{\arraystretch}{1.2}
\begin{tabular}{|c | c | l *{9}{c} c|}
\hline
\textbf{Feedback Round} &\textbf{Total Code Length} & \textbf{Configuration} 
& \textbf{R0} & \textbf{R1} & \textbf{REP} & \textbf{REP-2} & \textbf{PCR} & \textbf{SPC} & \textbf{SPC-2} & \textbf{RPC} & \textbf{LEAF} & \textbf{Total} \\
\hline

\multirow{2}{*}{$\mathbf{1^{st}}$} 
& \multirow{2}{*}{\textbf{3072}}
& Proposed nodes 
& 9 & 12 & 67 & 3 & 6 & 49 & 1 & 8 & 0 & 155 \\
& & Previous work  
& 123 & 63 & 48 & 1 & 4 & 52 & 1 & 1 & 69 & 362 \\
\hline

\multirow{2}{*}{$\mathbf{2^{nd}}$} 
&\multirow{2}{*}{\textbf{4096}} 
& Proposed nodes 
& 12 & 12 & 70 & 2 & 7 & 47 & 2 & 8 & 0 & 160 \\
& & Previous work 
& 166 & 63 & 45 & 0 & 4 & 51 & 3 & 3 & 89 & 424 \\
\hline

\multirow{2}{*}{$\mathbf{3^{rd}}$} 
&\multirow{2}{*}{\textbf{5120}} & Proposed nodes & 24 & 11 & 56 & 5 & 8 & 43 & 5 & 7 & 0 & 159 \\
& & Previous work 
& 213 & 59 & 37 & 2 & 4 & 50 & 3 & 2 & 101 & 471 \\
\hline

\multirow{2}{*}{$\mathbf{4^{th}}$} 
&\multirow{2}{*}{\textbf{6144}} 
& Proposed nodes 
& 23 & 10 & 57 & 5 & 8 & 44 & 5 & 7 & 0 & 159 \\
& & Previous work 
& 209 & 64 & 37 & 2 & 3 & 45 & 3 & 3 & 103 & 469 \\
\hline

\multirow{2}{*}{$\mathbf{5^{th}}$} 
&\multirow{2}{*}{\textbf{7168}} 
& Proposed nodes 
& 23 & 10 & 57 & 5 & 8 & 44 & 5 & 7 & 0 & 159 \\
& & Previous work  
& 209 & 64 & 37 & 2 & 3 & 45 & 3 & 3 & 103 & 469 \\
\hline

\multirow{2}{*}{$\mathbf{6^{th}}$} 
&\multirow{2}{*}{\textbf{8192}} 
& Proposed nodes 
& 23 & 10 & 57 & 5 & 8 & 44 & 5 & 7 & 0 & 159 \\
& & Previous work  
& 209 & 64 & 37 & 2 & 3 & 45 & 3 & 3 & 103 & 469 \\
\hline
\end{tabular}}
\vspace{-0.4cm}
\end{table*}
\section{Performance Evaluation}
 Fig. \ref{fig:FER_2} presents the decoding performance of each of the proposed configurations in an additive white Gaussian noise channel with Binary Phase Shift Keying modulation. In each of the feedback configurations, initially $2048$ bits are transmitted followed by 1024 bits transmitted in each additional retransmission round. In all cases, a $24$ bit CRC is used. Compared to the fast nodes scheme from \cite{sarkis_fast_2014,10293973}, we can see that our modifications to the special nodes that enable fast IR-HARQ do not result in any performance degradation. 
 Additionally, \autoref{tab:performance} shows the number of special node traversals in the two different configurations of the decoder used. The first is the proposed IR-HARQ and special node modifications as discussed in this work, and the second is only the special node modifications as presented in \cite{sarkis_fast_2014,10293973}. Without the proposed special nodes, the number of node traversals increases by up to $66\%$, impacting both throughput and latency. An example of the effect of the reduction of node traversals on the latency and throughput was shown in \cite{sarkis_fast_2014}. Introducing SPC nodes alone in \cite{sarkis_fast_2014} reduced the total node traversals by $58\%$, from $4915$ nodes (including R0, R1, and all nodes of size 4) to $2065$ nodes. This resulted in a $36\%$ reduction in decoder latency, from $5286$ clock cycles to $3360$ clock cycles. When all special nodes from \cite{sarkis_fast_2014} were employed, the resulting $71\%$ reduction in node traversals translated into a $19\times$ to $40\times$ throughput improvement and a $46\%$ decrease in latency, compared to a baseline architecture without the special nodes proposed in \cite{sarkis_fast_2014}. {\color{black} Assuming a similar hardware architecture as \cite{sarkis_fast_2014}, the proposed method is expected to yield latency reductions consistent with those reported therein, since the node structure is largely preserved. The added hardware in each special node consists of stages of 2-input XOR gates arranged in parallel. For REP nodes, a stage of $N_v$ XOR gates in parallel flip the sign bit of the LLRs being added, by taking the encoded \pcfrozen bits as their first input and the LLR sign bit as their second input and $N_v$ XOR gates in parallel are then used to generate the $\beta$. For SPC/SPC-2 and RPC nodes, a stage of $1$, $2$, and $3$ XOR gates respectively XOR the parities of the hard decisions with the \pcfrozen bits. From this structure, the added computational complexity amounts to $2 \times N_v$ XOR gates for REP, REP-2, and PCR nodes, and $1$, $2$, and $3$ XOR gates for SPC, SPC-2, and RPC nodes respectively. The corresponding path delay overhead is $2$ XOR gates for REP/REP-2/PCR nodes and $1$ XOR gate for SPC/SPC-2/RPC nodes, while the critical path of the decoder remains determined by the slowest node, which was the SPC node in \cite{sarkis_fast_2014}.}
 
 For $N_v=128$ with 6 quantization bits, we estimate the hardware complexity using the NAND gate equivalent approximation method described in \cite{jalaleddine_hardware-friendly_2025}. Under this model, the REP node requires $10958$ NAND gates, while the SPC node requires $6618$ NAND gates. {\color{black} The proposed IR-HARQ modifications introduce $1024$ additional NAND gates for the REP node and $4$ for the SPC node, representing less than $9.4\%$ and $0.06\%$ overhead, respectively. Additionally, $\frac{N_v}{2}\log_2(N_v)$ 2-input XOR gates are needed to encode the \pcfrozen bits, which can be done in advance without impacting the path delay. Hence, a total of $1792$ NAND gates are required to generate the encoded \pcfrozen input of all the special nodes.} Furthermore, the hardware designed for the REP node can be reused for the REP-2 and PCR nodes, and similarly, the SPC node hardware can be adapted for the SPC-2 and RPC nodes with only minor modifications.
\section{Conclusion}
In this work, we present a series of modifications to the state-of-the-art special nodes to enable the IR-HARQ support. Our simulation results verify that modifications to the existing R0, R1, REP, REP 2, SPC, SPC 2, RPC and PCR nodes enable the use of special nodes in the IR-HARQ scheme without any decoding performance degradation compared to the scheme that uses traditional nodes. {Additionally, our proposed modifications allow us to reduce the number of node traversals with polar codes by up to $66\%$ compared to the IR-HARQ scheme utilizing the unmodified nodes. This enables }IR-HARQ support in future standards for polar SC { with little hardware complexity overhead}.
\bibliographystyle{ieeetr}
\bibliography{IEEEabrv,new}
\end{document}